# A thermal beam of metastable krypton atoms produced by optical excitation

Y. Ding,[1,2,a)] S.-M. Hu,[3] K. Bailey,[2] A. M. Davis,[1,4] R. W. Dunford,[5] Z.-T. Lu,[1,2,6], T. P. O'Connor,[2] L. Young,[5]

[1]*Enrico Fermi Institute, The University of Chicago, Chicago, IL 60637*
[2]*Physics Division, Argonne National Laboratory, Argonne, Illinois 60439*
[3]*Hefei National Laboratory for Physical Sciences at the Microscale, University of Science and Technology of China, Hefei, 230026, China*
[4]*Department of the Geophysical Sciences, The University of Chicago, Chicago, IL 60637*
[5]*Chemistry Division, Argonne National Laboratory, Argonne, Illinois 60439*
[6]*Department of Physics, The University of Chicago, Chicago, IL 60637*

A room-temperature beam of krypton atoms in the metastable $5s[3/2]_2$ level is demonstrated via an optical excitation method. A Kr-discharge lamp is used to produce VUV photons at 124 nm for the first-step excitation from the ground-level $4p^6\ ^1S_0$ to the $5s[3/2]_1$ level. An 819 nm Ti:Sapphire laser is used for the second-step excitation from $5s[3/2]_1$ to $5p[3/2]_2$, followed by a spontaneous decay to the $5s[3/2]_2$ metastable level. A metastable atomic beam with an angular flux density of $3 \times 10^{14}$ s$^{-1}$sr$^{-1}$ is achieved at the total gas flow rate of 0.01 cm$^3$ STP /s (or $3 \times 10^{17}$ atoms/s). The dependences of the flux on the gas flow rate, laser power, and lamp parameters are investigated.

a) Corresponding author: ding@anl.gov

I. Introduction

Thermal beams of metastable noble gas atoms have a wide range of applications including atom lithography,[1] atom holography,[2] atom optics,[3] cold atomic collision studies,[4] and precision measurements.[5] Sources of such beams have been developed based on electron-impact excitation where energetic electrons are produced in various ways such as an electron beam,[6] a DC glow discharge,[7] and a surface-wave-sustained plasma.[8] These sources produce intense metastable atomic beams with angular flux densities in the range of $10^{13}$-$10^{15}$ $s^{-1}sr^{-1}$. The gas flow or consumption rate, however, had not been a primary design concern in prior publications because the aforementioned applications use gases that are readily available in large quantities (> 1000 $cm^3$ STP).

Atom Trap Trace Analysis,[9] in which laser manipulation of metastable krypton (Kr*) atoms is used to count atoms of rare isotopes, has motivated the development of an efficient source of Kr* atoms that can accommodate a small amount of gas (< 0.1 $cm^3$ STP). Indeed, ATTA demands both high beam flux and high excitation efficiency. For this application, Chen et al. developed a thermal beam of Kr* atoms at the 5s[3/2]$_2$ level extracted from a RF-driven discharge,[10] and achieved an angular flux density of $4 \times 10^{14}$ $s^{-1}sr^{-1}$ at the gas flow rate of $7 \times 10^{16}$ Kr atoms/s. The excitation efficiency, i.e. the ratio of the metastable Kr* atom flux over the ground-level Kr atom flux in the forward direction, was approximately $10^{-3}$. This efficiency seems to have been limited by the inherent ionization process present in the discharge. It was observed that increasing the discharge intensity above an optimum level resulted in a lower Kr* beam flux. More recently, Ding et al. modified the source by replacing the RF-driven discharge with a microwave-driven discharge, and achieved comparable results.[11]

A drastically different approach is to transfer the atoms from the ground-level to the metastable level via a multi-photon excitation process. In 2002, Young et al. developed an optical excitation scheme to produce metastable Kr atoms and demonstrated its application in a gas cell.[12] Here we report the demonstration of a thermal Kr* beam based on the same optical excitation scheme. With further developments, this type of source can potentially produce a Kr* beam with both higher flux and higher excitation efficiency than a conventional discharge source.

Figure 1 shows the relevant atomic energy levels of Kr. Due to selection rules on both angular momentum and parity, the direct transition between the ground-level $4p^6\ ^1S_0$ and $5s[3/2]_2$ is forbidden, causing $5s[3/2]_2$ to be metastable with a lifetime of 40 s. Consequently, once populated in the $5s[3/2]_2$ level, the metastable Kr* atoms can be trapped and cooled with laser light at 811 nm resonant to the cycling transition $5s[3/2]_2$ - $5p[3/2]_3$. The path from $4p^6\ ^1S_0$ to $5s[3/2]_2$ requires a minimum of three allowed transitions. In the scheme shown by Young *et al.*, the atom is first excited from the ground-level $4p^6\ ^1S_0$ to $5s[3/2]_1$ with a VUV photon at 124 nm emitted from a Kr-discharge lamp;[13,14] the atom is then excited from $5s[3/2]_1$ to $5p[3/2]_2$ with an infrared photon from an 819 nm laser; the atom then spontaneously decays to $5s[3/2]_2$ with a branching ratio of 90%, thus completing the transfer. Besides this two-step (VUV + IR) excitation scheme, $5p[3/2]_2$ can also be reached via a two-photon (UV + UV) transition induced by a single laser at 215 nm. However, present UV lasers, pulsed or cw, are not yet powerful enough for this latter scheme to be competitive.

**II. Experiment and Discussion**

A schematic of the atomic beam apparatus is shown in figure 2. Krypton gas flows into the excitation chamber through a capillary-plate collimator that has a diameter of 13 mm, a thickness of 2 mm, and a closely packed matrix of 0.010 mm diameter holes. The chambers are differentially pumped by two turbopumps. The typical pressures are $10^{-4}$ Torr in the excitation chamber and $10^{-5}$ Torr in the detection chamber. The first-step excitation is induced by VUV photons emitted from a Kr-discharge lamp with a $MgF_2$ front window (diameter ~ 10 mm), which is placed within 1 cm from the atomic beam. The intensity of the VUV photons is measured using a XUV photodiode detector which is fitted with a filter with a ~16% peak transmission at 121.8 nm and a width of 12.75 nm. We measure that the lamp typically emits $3 \times 10^{16}$ VUV photons $s^{-1}sr^{-1}$ with a 100 Watts of microwave power. The second-step excitation is induced by infrared photons (819 nm) from a ring Ti:Sapphire laser. The typical laser power is 300 mW and a retro-reflecting mirror doubles the power to 600 mW at the excitation region. The diameter of the laser beam at the excitation region is approximately 15 mm, comparable to the sizes of the atomic beam collimator and the lamp window.

The Kr* beam flux is measured in the detection chamber by laser induced fluorescence using a diode laser whose wavelength is tuned to the resonance of the cycling transition $5s[3/2]_2 - 5p[5/2]_3$ at 811.5 nm. Two laser beams cross the atomic beam at the center of the detection chamber, which is 33 cm away from the capillary plate (Fig. 2). The resonance fluorescence of the atoms is monitored through a lock-in amplifier using a Si photodiode fitted with a bandpass filter. The filter has a bandpass of 10 nm and provide a ~50% peak transmission at 810 nm. Figure 3 shows a typical fluorescence signal as the probe 811 nm laser is scanned across the resonances. One laser beam, crosses perpendicular to the atomic beam, provides the Doppler-reduced fluorescence signal that appears as a narrow peak. From the intensity and the known transition rates, we calculate the total atomic beam flux. The other laser beam, crosses at a 45º angle to the atomic beam, produces a Doppler-shifted and broadened fluorescence signal. From this we calculate that the most probable velocity in the Kr* beam is 280 m/s and that the Kr* beam is at room temperature.

Figure 4 shows the Kr* beam flux dependence on the gas flow rate, which is proportional to the pressure in the excitation chamber. At low pressure, the Kr* beam flux increases with the gas flow rate and chamber pressure. The flux saturates at a chamber pressure of $1 \times 10^{-4}$ Torr. At this pressure the mean free path of the Kr* atoms becomes comparable to the traveling distance of approximately 15 cm; and the loss of Kr* atoms due to collisions with background atoms starts to take effect. Above $4 \times 10^{-4}$ Torr, the collisional loss becomes dominant and the Kr* beam flux drops dramatically with increasing pressure.

Figure 5 shows the dependence of the Kr* beam flux on the power of the 819 nm laser beam for the second-step excitation. The laser beam has a diameter of approximately 15 mm, and is retro-reflected to double the power in the excitation region.

The power cited here and in the figure caption is the total power including the retro-reflected beam. The Kr* beam flux increases with the laser beam power. The second-step transition is nearly saturated at a power of 700 mW and an intensity of 400 mW/cm$^2$.

The power and bandwidth of the VUV output of the Kr-discharge lamp is the critical element for source performance. Figure 6 shows the dependence of the Kr* beam flux on the Kr partial pressure in the lamp. We note that the Kr gas in the lamp is completely separated from the Kr gas in the atomic beam. In an actual application, the Kr atoms in the atomic beam are extracted from precious geological samples, while regular, commercially available Kr gas is adequate to fill the lamp. The lamp was filled with either pure Kr gas or a mixture of Kr and He (Kr/He = 10%). In the case of pure Kr gas, the Kr* beam flux increases with decreasing of the Kr pressure in the lamp (Fig. 6). We believe that this dependence is due to the interplay between the photon emission from the excited krypton atoms in the lamp and the resonant absorption of photons by the ground-level krypton atoms also in the lamp. At high lamp pressure (> 200 mTorr), the self-absorption effect[15] becomes dominant, resulting in a low Kr* beam flux. Lower lamp pressures are preferred. However, when the lamp pressure is below 50 mTorr, it becomes difficult to drive the discharge towards the front window which is 10 cm away from the microwave cavity. In this case, the Kr* flux drops dramatically (Fig. 6) since the optical depth for the VUV emission is on the order of 10$^{-6}$ m at this pressure. In the case of Kr-He gas mixture, the helium gas helps to maintain the discharge, allowing the Kr partial pressure to be lowered to 30 mTorr, and thereby achieving a higher Kr* beam flux (Fig. 6). We believe that by moving the microwave cavity closer to the front window the discharge can be maintained near the front window at an even lower pressure, and the Kr* beam flux can be significantly increased. However, this idea has not yet been tested. A measurement performed by J. Reader on a similar lamp run at a mode without self-reversal (no "hole" in the middle of the line profile) showed that the width of this VUV emission line was approximately 300 GHz, much larger than the Doppler broadened absorption width and isotopic shifts.[16] We have performed a separate test on the spectrum purity of the lamp. An interference filter with a ~16% peak transmission at 121.8 nm and a width of 12.75 nm was installed in front of the lamp in order to cut off visible and infrared photons that might pump Kr* atoms out of the metastable level. The Kr* beam flux only decreased by a factor approximately equal to the transmission of the filter at 124 nm.

When all the aforementioned conditions are optimized, we achieved a $^{84}$Kr* beam flux density of $2 \times 10^{14}$ s$^{-1}$ sr$^{-1}$, which is equivalent to an atomic beam flux density of $3 \times 10^{14}$ s$^{-1}$ sr$^{-1}$ at 100% isotopic abundance. The maximum flux is achieved at a total gas flow rate of $3 \times 10^{17}$ Kr atoms/s (or $1 \times 10^{-2}$ cm$^3$ STP/s). The relative uncertainty of the flux value is 50%, which derives mainly from the uncertainty in estimating the fluorescing volume. Compared to a source based on a RF-driven discharge,[10] the Kr* beam flux density achieved here is nearly equal, however, the excitation efficiency is five times lower because of the higher gas flow rate.

In a discharge-based source, atoms of all isotopes are transferred to the metastable level with equal probability. In the optical excitation scheme, however, the second-step excitation by the 819 nm laser is isotopically selective due to isotope shifts. When it is tuned to the resonance of $^{84}$Kr (isotopic abundance = 57%), the excitation of $^{86}$Kr (17%) atoms is suppressed as shown in Fig. 3. On the other hand, by tuning the laser to the resonances of $^{86}$Kr (17%) and $^{82}$Kr (12%), respectively, we have measured the Kr* beam flux of these two isotopes, and found the ratios of fluxes to be consistent with the ratios of isotopic abundances. A unique case is $^{83}$Kr (11%, I = 9/2), which possesses hyperfine structure. Among those transitions from 5s[3/2]$_1$ to 5p[3/2]$_2$ at 819 nm, the strongest one is F = 11/2 → F' = 13/2. Assuming the transition is saturated, we calculate that the total branching ratio for the transfer is 32% and, when combined with the ratio of isotopic abundances of 11%, the ratio of fluxes $^{83}$Kr/$^{84}$Kr is expected to be 0.06. Interestingly, the ratio of fluxes $^{83}$Kr/$^{84}$Kr was actually measured to be 0.15, approximately two times higher than expected and just slightly less than the ratio of isotopic abundances. The additional flux of $^{83}$Kr may come from other transitions not included in the calculation, or be due to better lamp performance at the $^{83}$Kr lines.

**III. Outlook**

For ATTA applications, this new type of source offers several important advantages over the traditional sources based on gas discharges. In a discharge-based Kr* beam source, a small fraction of the Kr atoms are continuously ionized and imbedded into walls, meanwhile, some Kr atoms are removed from the surface of the walls. This effect causes a slow but continuous exchange of Kr atoms between the sample and the walls, resulting in loss of sample and, worse, cross-sample contamination.[17] This problem is avoided in an optical excitation source. Moreover, a discharge-based source requires a minimum operation pressure of a few mTorr. An optical excitation source can operate at arbitrarily low pressure, and is only limited by the statistics of atom counting. Although the excitation efficiency achieved with this source is five times lower than that of a RF-driven discharge source, we see potential improvements, particularly on the lamp design, which can be implemented in later versions. We also note that the on-going development of VUV lasers at 124 nm is of considerable interest.[18, 19]

**Acknowledgement**

We thank W. Wong, X. Du, and R. Matsuoka for their contributions to the development of the setup at various stages of this project. We also thank S. J. Freedman, J. R. Guest, R. J. Holt, P. Mueller, and J. Reader for helpful discussions. This work was supported by the U.S. Department of Energy, Office of Nuclear Physics and Office of Basic Energy Sciences, under Contract No. W-31-109-ENG-38, and by a University of Chicago-Argonne National Laboratory collaborative seed grant.

# Figure Captions

**Figure 1**. Kr atomic energy level diagram. The $5s[3/2]_2$ level is metastable with a lifetime of 40 s. The $5s[3/2]_2$ - $5p[3/2]_3$ transition at 811 nm is used for laser trapping and detection of the metastable Kr* atoms. The first-step excitation, $4p^6\ ^1S_0$ - $5s[3/2]_1$, is induced by 124 nm photons from a Kr-discharge lamp. The second-step excitation, $5s[3/2]_1$ - $5p[3/2]_2$, is induced by 819 nm photons from a Ti:Sapphire laser. The third step of the transfer process is a spontaneous decay from $5p[3/2]_2$ to $5s[3/2]_2$.

**Figure 2.** Schematic of the experimental apparatus.

**Figure 3.** $^{84}$Kr* beam fluorescence vs. laser frequency. The narrow peak is due to the Doppler-reduced fluorescence induced by the perpendicular laser beam. The broad peak is due to the Doppler-broadened fluorescence induced by the 45 degree laser beam. The continuous curve is a fit of data with Maxwell-Boltzmann distribution at room temperature. The $^{86}$Kr* peak is weaker than expected from its isotopic abundance ($^{86}$Kr/$^{84}$Kr = 0.3) due to the fact that the second-step excitation by the 819 nm laser is isotopically selective.

**Figure 4.** $^{84}$Kr* beam flux vs. excitation chamber pressure. The pressure in the excitation chamber is proportional to the krypton gas flow rate. The flux first increases linearly with the gas flow rate at low pressure, then saturates and decreases due to increased collisional loss in the chamber.

**Figure 5.** $^{84}$Kr* beam flux vs. 819 nm laser power. The second-step transition is nearly saturated at 700 mW.

**Figure 6.** $^{84}$Kr* beam flux vs. Kr partial pressure in the lamp. The solid squares are data recorded with a pure Kr gas lamp; the open circles are data with a gas mixture (Kr/He = 10%) lamp. At very low pressure (solid triangle), the discharge becomes unstable and the flux drops dramatically. The dashed line, based on Kr pressure inverse proportion, is drawn to guide the eye.

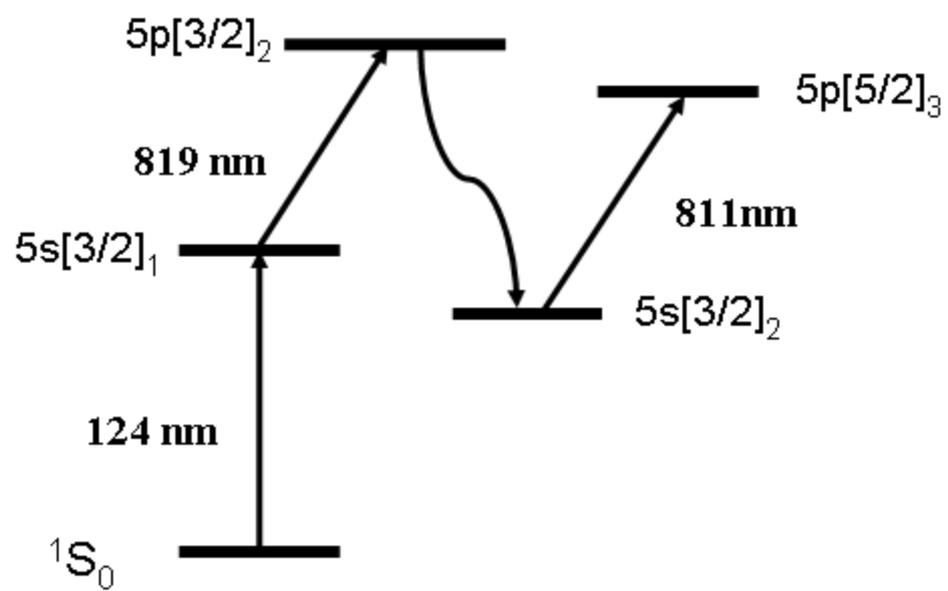

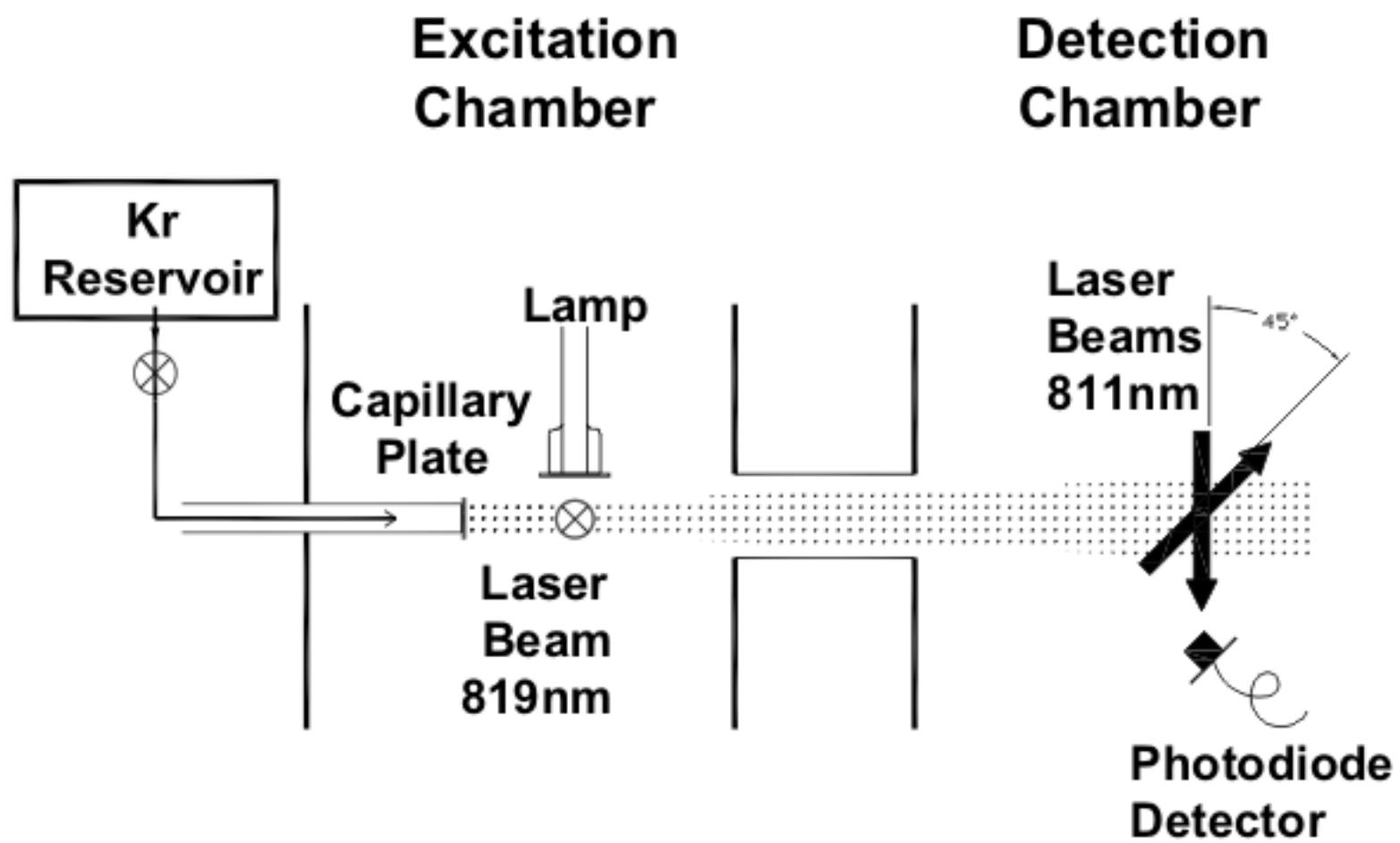

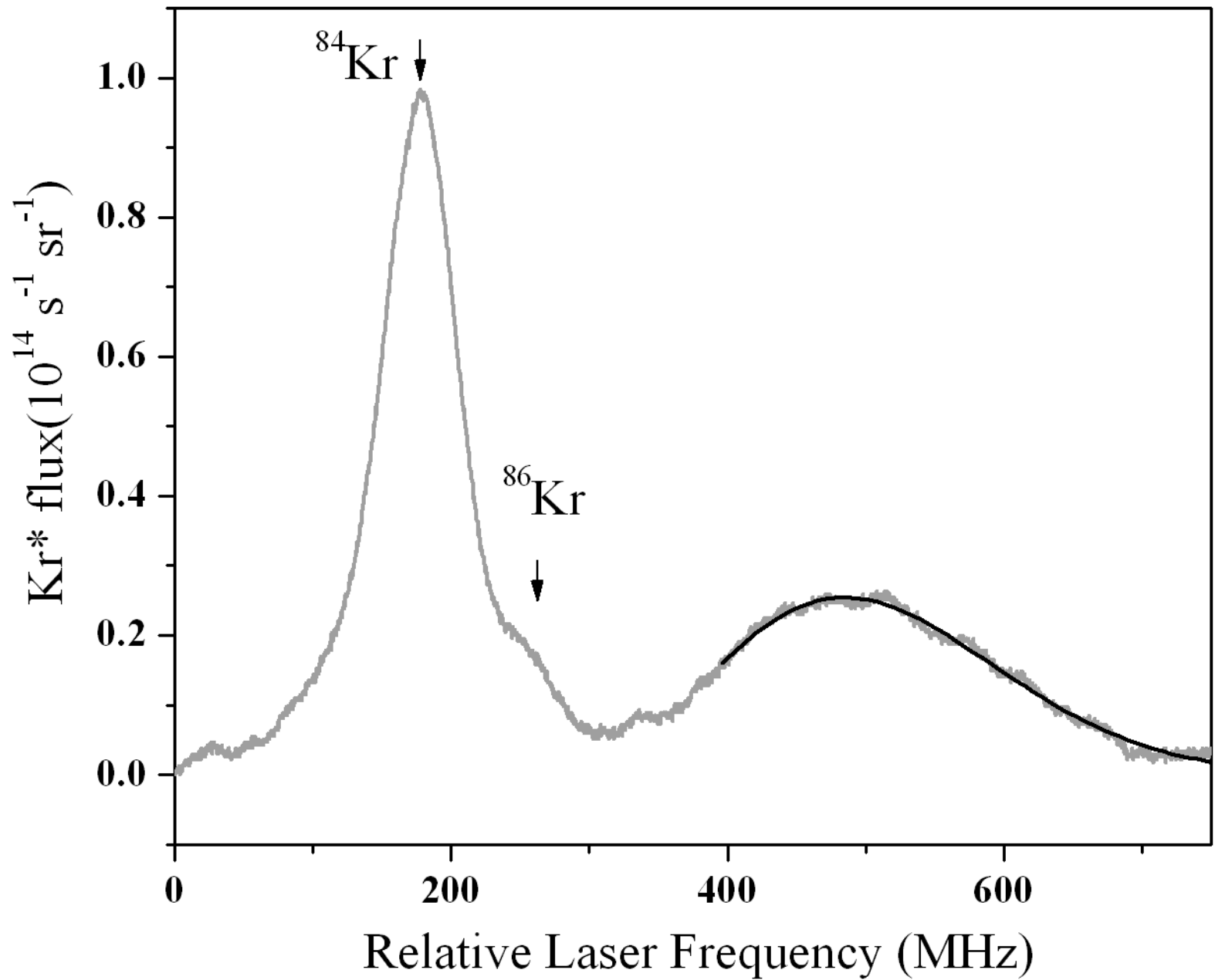

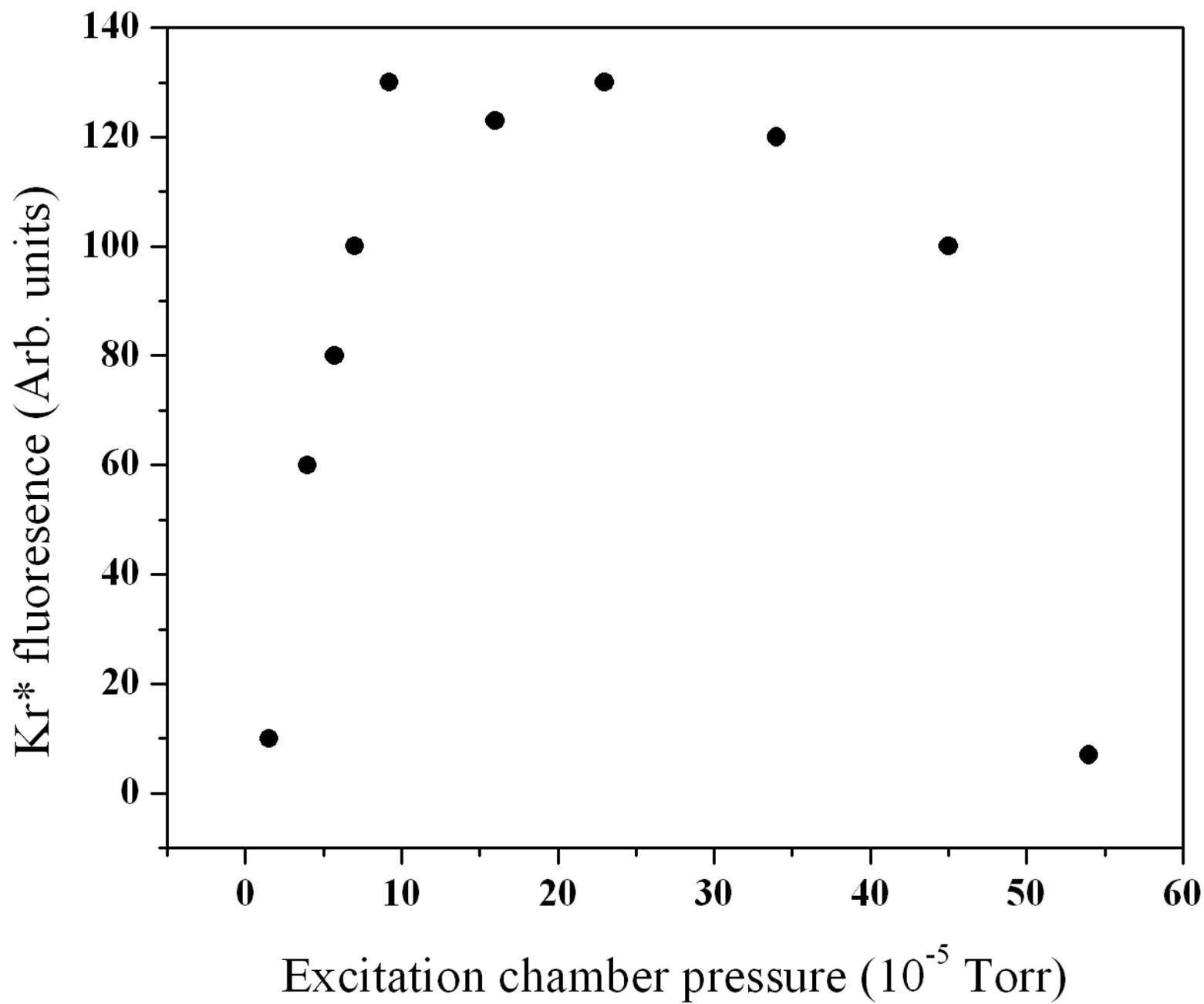

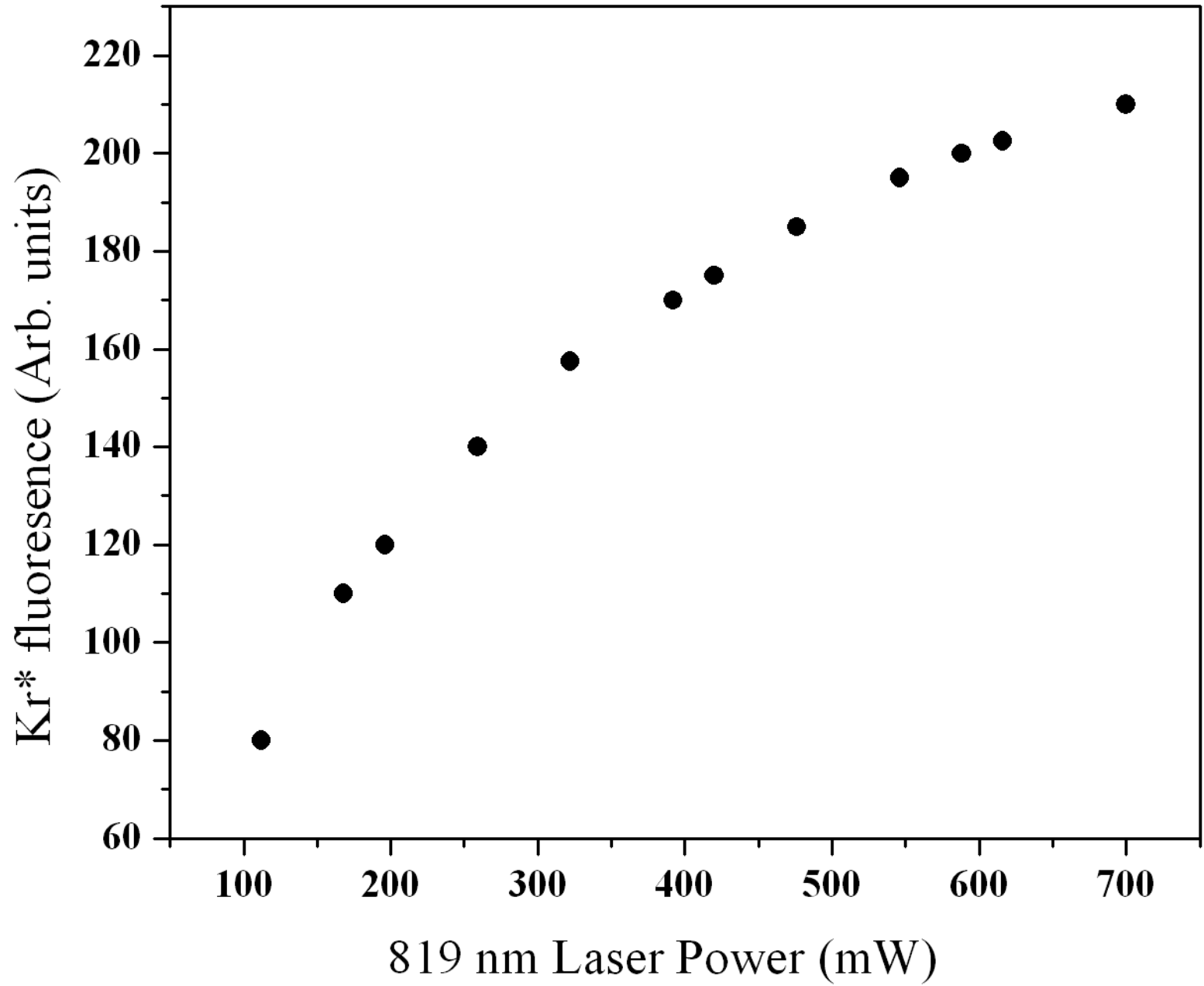

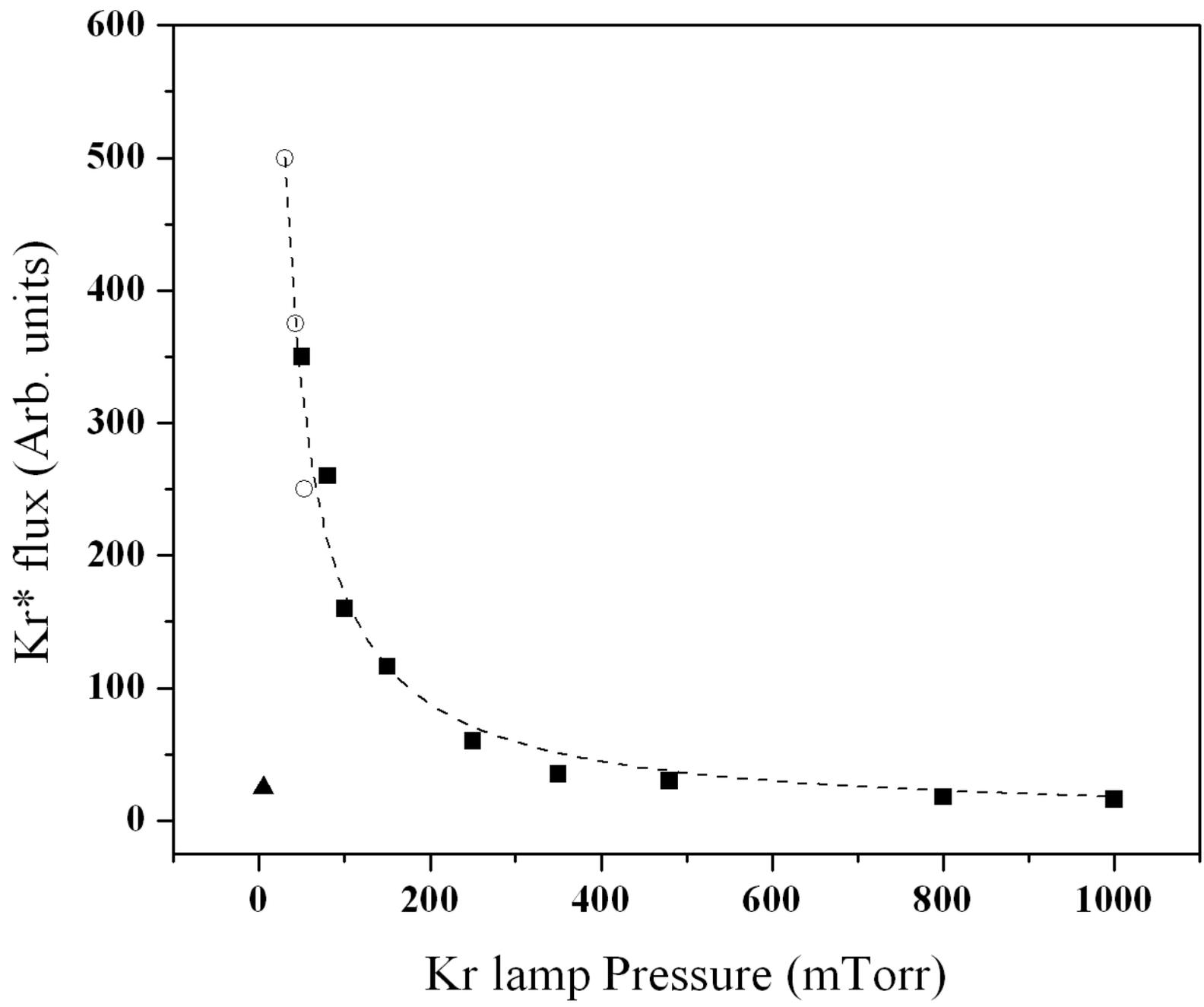